\documentstyle[prl,aps,floats,epsf,twocolumn]{revtex}
\catcode`\@=11
\def\simle{\mathrel{\mathpalette\@versim<}}   
\def\simge{\mathrel{\mathpalette\@versim>}}   
\def\@versim#1#2{\lower2.5pt\vbox{\baselineskip0pt \lineskip-.5pt
   \ialign{$\m@th#1\hfil##\hfil$\crcr#2\crcr\sim\crcr}}}
\begin{document}

\draft
\twocolumn[\hsize\textwidth\columnwidth\hsize\csname @twocolumnfalse\endcsname

\title{
Re-entrant Behavior and Gigantic Response in Disordered Spin-Peierls System  
}
\author{
Hitoshi Seo$^{1,2},$Yukitoshi Motome$^3$, and Naoto Nagaosa$^{1,4,5}$
}
\address{
$^{1}$Correlated Electron Research Center (CERC), AIST Tsukuba Central 4, 
Ibaraki 305-8562, Japan
}
\address{
$^{2}$Domestic Research Fellow, JSPS, Chiyoda-ku, Tokyo 102-8471, Japan
}
\address{
$^{3}$Condensed-Matter Theory Laboratory, RIKEN, Wako, Saitama 351-0198, Japan
}
\address{
$^{4}$Tokura Spin SuperStructure Project (SSS), ERATO, JST,
c/o AIST Tsukuba Central 4, Ibaraki 305-8562, Japan
}
\address{
$^{5}$CREST, Department of Applied Physics, University of Tokyo, 
Bunkyo-ku, Tokyo 113-8656, Japan
}
\date{\today}

\maketitle

\begin{abstract}
Effects of disorder and external field 
on the competing spin-Peierls  
and antiferromagnetic states are studied theoretically 
in terms of the numerical transfer matrix method applied  
to a quasi one-dimensional spin 1/2 Heisenberg model 
coupled to the lattice degree of freedom. 
We show that, at temperatures above 
the impurity-induced antiferromagnetic phase, 
inhomogeneous spin-Peierls lattice distortions remain to exist 
showing a re-entrant behavior. 
This feature can be drastically altered 
by very weak perturbations, e.g., the staggered magnetic 
field or the change in interchain exchange coupling 
$J_\perp$, leading to a huge response, 
which is analogous to the colossal magnetoresistance phenomenon 
in perovskite manganese oxides. 
\end{abstract}

]


In strongly correlated electronic systems, 
various kinds of long ranged ordered (LRO) phases emerge, 
frequently those energies being very close to each other, 
i.e., with multicriticality. 
In such cases, the system becomes sensitive to perturbations 
such as applied pressure, magnetic field, electric field,  
carrier/impurity doping, etc., 
leading to intriguing new phenomena. 

\begin{figure}
\epsfxsize=7cm
\centerline{\epsfbox{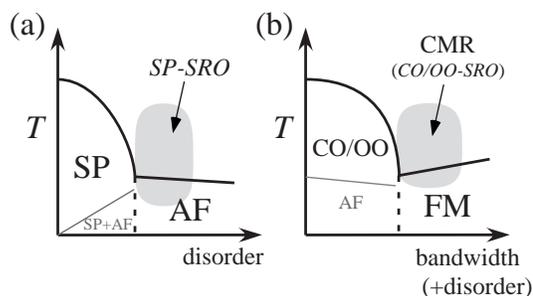}}
\caption{Schematic experimental phase diagrams for 
(a) disordered SP system~\protect\cite{Uchinokura} 
and (b) manganite oxides showing pronounced CMR~\protect\cite{Tomioka55}. 
The ordinate is temperature, $T$, 
while the abscissa represents
(a) the degree of disorder realized by increasing the impurity 
concentration 
and (b) the substituent concentration controlling 
not only the degree of disorder but also the bandwidth 
(for example Ca concentration in 
Sm$_{0.55}$(Ca,Sr)$_{0.45}$MnO$_3$~\protect\cite{Tomioka55}). 
The grey areas show (a) the region of re-entrant SRO of SP state,  
and (b) the region with enhanced CO/OO fluctuation 
and CMR effect. 
}
\label{analogy}
\end{figure}

The disordered spin-Peierls (SP) system 
realized in doped CuGeO$_3$~\cite{Hase} 
is a typical example 
showing such subtle multicriticality 
of different phases. 
Its phase diagram 
on the plane of temperature, $T$, and impurity concentration, $x$,  
is experimentally explored in detail (Fig.~\ref{analogy}(a)), 
which is almost universal regardless of the dopant
substituting for either Cu or Ge sites~\cite{Uchinokura}. 
The nonmagnetic SP state realized in undoped CuGeO$_3$ 
is fragilely destabilized by impurities of a few percent, 
as shown in Fig.~\ref{analogy}(a). 
The impurities induce local spin moments 
where the interchain exchange coupling 
results in the antiferromagnetic (AF) state, 
which, at the small-$x$ and low-$T$ region, 
co-exists with SP lattice distortion 
in an inhomogeneous way 
(SP+AF phase in Fig.~\ref{analogy}(a))~\cite{Regnault}. 
On the other hand, when $x$ is rather larger, 
a uniform AF state with 
homogeneous staggered spin moments becomes stabilized.
The critical region between the phases 
is slightly sensitive to the dopant, 
where the system shows either bicritical behavior 
between the SP and AF states, 
or tetracritical behavior between all the three
phases~\cite{Uchinokura}, 
though these are not thoroughly traced by experiments. 
Fig.~\ref{analogy}(a) shows the former case with 
a bicritical point. 
Theoretical aspects of this system 
are also intensively studied 
analytically~\cite{Fukuyama,Khomskii,Fabrizio,Mostovoy,Saito} 
as well as numerically~\cite{Poilblanc}, 
mostly to understand the novel co-existent state 
at the small-$x$ region, 
while a few mention the large-$x$ region~\cite{Mostovoy,Saito}. 

Another typical example of such interplay between 
multicriticality and disorder 
in strongly correlated electron systems 
is the colossal magnetoresistance (CMR) 
in perovskite manganese oxides 
$R_{1-x}A_{x}$MnO$_3$ ($R$: trivalent rare earth element, 
$A$: divalent alkaline earth element)~\cite{CMRreview}.  
CMR is seen in compounds 
where an insulator-to-ferromagnetic metal (FM)
phase transition is observed upon cooling. 
Small magnetic field can 
make the characteristic temperature 
such as the maximum in resistivity drastically shifted 
to higher temperature, 
thus resulting in a ``colossal'' change of the resistivity.  
To understand this effect, 
the importance of 
not only the multicriticality 
between the FM phase and 
the insulating charge/orbital ordered (CO/OO)
state~\cite{TomiokaPCMO,Murakami},
but also the disorder effect 
due to quenched chemical dopant 
has recently been pointed out~\cite{Mn_disorder,Tomioka55}. 
From experiments carefully controlling 
the degree of disorder in a chemical way~\cite{Tomioka55}, 
the randomness seems to 
destabilize the CO/OO state resulting in 
the FM ground state~\cite{note_glassy} 
where the CMR effect becomes prominent, 
as schematically shown in Fig.~\ref{analogy}(b).
Theoretical works on models for manganites with randomness 
have succeeded to reproduce such features~\cite{Motome,Aliaga}. 

Here we point out an analogy between 
the CMR manganites and the disordered SP system 
by making a correspondence of the competing phases as 
CO/OO $\leftrightarrow$ SP and 
FM $\leftrightarrow$ AF, respectively.  
As discussed above, the former states are 
destabilized by impurities, resulting in the latter states. 
As a matter of fact, 
their phase diagrams on the plane of $T$ and 
degree of disorder, as seen in Fig.~\ref{analogy}, 
are isomorphic to each other. 
In the disordered SP system, 
above the transition temperature 
for the uniform AF state, $T_{\rm AF}$, 
short range order (SRO) of SP lattice distortion 
is observed to develop toward $T_{\rm AF}$ 
while it is suppressed below it, 
resulting in a characteristic re-entrant behavior~\cite{Nakao}. 
This is also analogous to the CMR compounds 
where SRO of lattice modulations due to CO/OO, 
either static or dynamic, 
are seen above the FM transition temperature~\cite{CMRreview,Tomioka55}. 
The origin of the CMR effect is now believed to be due to 
the sensitivity of such re-entrant 
CO/OO fluctuations~\cite{Mn_disorder,Tomioka55,Motome,Aliaga}. 
Then, similarly, 
we expect that some ``colossal'' effect 
can also be seen in disordered SP system, 
which is investigated in this Letter. 

To study the disordered SP system, 
we consider a quasi 1D spin 1/2 AF Heisenberg model  
coupled to the lattice distortions~\cite{SPreview,Inagaki}. 
The model consists of 1D chains, each of which being 
given by the Hamiltonian, 
\begin{equation}
{\cal H_{\rm 1D}}=\sum_{i} \left\{
 J (1+u_{i}) {\vec S}_{i} \cdot {\vec S}_{i+1} 
 + \frac{K_{i}}{2} u_{i}^2 \right\} \ ,
\label{H_SP}
\end{equation}
while these are coupled by interchain exchange interaction 
between neighboring chains, 
\begin{equation}
{\cal H_{\perp}}=\sum_{\langle i,j \rangle} 
J_{\perp} {\vec S}_{i} \cdot {\vec S}_{j} \ ,
\label{H_perp}
\end{equation}
where $J,J_{\perp} > 0$. 
Each chain is the usual SP model 
if $K_{i}$ is uniform, which is 
unstable toward SP lattice dimerization, $u_i=(-1)^i u_{\rm SP}^0$, 
in order to gain singlet formation energy 
in spite of the loss in elastic energy 
in the second term of Eq.~(\ref{H_SP})\cite{SPreview}.

On the other hand, the interchain exchange coupling, $J_{\perp}$, 
in Eq.~(\ref{H_perp}) 
brings about the competing AF state~\cite{Inagaki}, 
to which we apply the mean field approximation 
as in Refs.~\cite{Saito,Inagaki}.
This is justified for quasi-1D system because the correlation 
length is already large near the transition temperature~\cite{Imry}. 
Then we obtain an effective 1D model, 
\begin{equation}
{\cal H'}={\cal H_{\rm 1D}} 
- z J_{\perp} \sum_i \left( \langle S^z_{i} \rangle S^z_{i} 
- \langle S^z_{i} \rangle^2 /2 \right),
\label{H_1D}
\end{equation}
where $z$ is the number of neighboring chains 
and the mean field approximation for $J_\perp$ is even exact
in the limit of $z \to \infty$ with $z J_{\perp}$ being
finite. Note that in this formalism 
the mean field $\langle S^z_i \rangle$ is independent of
chain in the original model, 
and thus consequently $u_i$ too~\cite{Saito}.

Impurities are modeled here by replacing the spin-lattice coupling 
constant, $K_i$, as $K_{\rm bulk} \rightarrow K_{\rm imp}$ 
randomly~\cite{Khomskii,Saito}, 
and the renormalized lattice distortions $u_i$ are treated as classical values. 
We use the numerical transfer matrix method~\cite{Betsuyaku,note} 
to calculate the finite-$T$ properties of this model, 
determining $u_i$ and $\langle S^z_i \rangle $ 
for each site on the chain self-consistently 
by minimizing the free energy. 
We use data on open chains with sizes up to $L=160$ 
and obtain thermodynamic-limit properties 
applying extrapolation by $1/L$, 
and for the random average 
we take 20-40 samples 
fixing the impurity concentration. 
Note that the method looses accuracy at low-$T$, 
at about typically $T<0.1J$, 
thus our study is complementary to the previous studies 
on the low-$T$ properties 
of disordered SP system~\cite{Fukuyama,Saito}. 
In the following we set $J=1$ as the energy unit. 

\begin{figure}
\epsfxsize=5.3cm
\centerline{\epsfbox{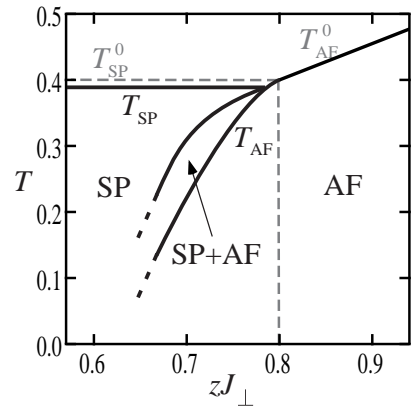}}
\caption{Theoretical phase diagram for 
disordered SP system on the plane of $T$ and $zJ_{\perp}$. 
The phase boundaries for the pure case are also written 
as grey broken lines. 
The AF transition temperature for the pure case, $T_{\rm AF}^0$, 
is identical to that in the presence of disorder, 
$T_{\rm AF}$ for $zJ_{\perp}>0.8$.
}
\label{phasediagram}
\end{figure}
In Fig.~\ref{phasediagram} we show the phase diagram 
on the plane of $T$ and $zJ_{\perp}$ 
for the case of fixed impurity concentration of 1/8 
with $K_{\rm bulk}=0.8$ and $K_{\rm imp}=1.5$. 
In the pure case of $K_{\rm bulk}=0.8$, 
drawn by grey broken lines in Fig.~\ref{phasediagram}, 
a bicritical behavior is seen at $zJ_{\perp} = 0.8$ 
between the SP state 
and the AF state with $\langle S^z_i \rangle =(-1)^i m_{\rm AF}^0$, 
$m_{\rm AF}^0$ being the staggered spin moment, 
naturally understood from the previous studies for the ground
state~\cite{Inagaki}. 
Here the SP transition temperature, $T_{\rm SP} ^0$, is constant 
and the AF transition temperature, $T_{\rm AF} ^0$, is linear 
as a function of $zJ_{\perp}$, 
since we treat $u_i$ as classical values 
and the interchain interaction within mean field approximation.
When the impurities are introduced, 
as can be seen in Fig.~\ref{phasediagram}, 
the AF state is extended over a certain range of $zJ_{\perp}$. 
However, above this uniform AF phase, 
we observe the SP state 
and a crossover region in between, 
where a co-existent state of SP and AF 
with both order parameters spatially varying 
is stabilized, represented as SP+AF in Fig.~\ref{phasediagram}. 
We shall see this in more detail in the following. 

\begin{figure}
\epsfxsize=6.5cm
\centerline{\epsfbox{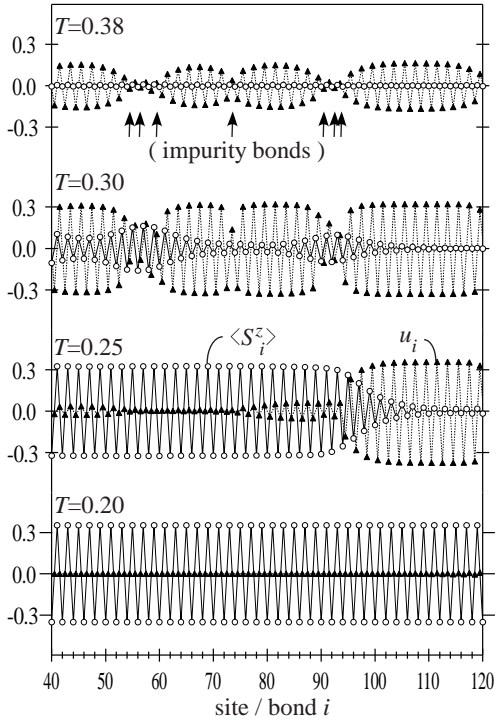}}
\caption{Spatial pattern of $u_i$ and $\langle S^z_i \rangle $
for a certain sample of $L=160$ with $zJ_{\perp}=0.72$. 
Only the middle half of the chain ($L/4 < i < 3L/4$) is shown, 
to omit open boundary effects. 
The impurity positions with $K_{\rm imp}=1.5$ 
(in the bulk, $K_{\rm bulk}=0.8$)
are shown by arrows. 
}
\label{realspace}
\end{figure}
As an example, in Fig.~\ref{realspace} 
the evolution of spatial pattern of $u_i$ and $\langle S^z_i \rangle$ 
is shown for a certain sample of $L=160$ for $zJ_{\perp}=0.72$. 
When $T$ is decreased from high temperatures, 
the SP lattice dimerization is developed from slightly below $T_{\rm SP} ^0$, 
but the amplitude is modulated such that  
$\left| u_i \right|$ show local minima at the impurity bonds, 
since $K_{\rm imp}>K_{\rm bulk}$ dislikes lattice distortion. 
When $T$ is lowered further, 
AF staggered moment starts to emerge, 
locally nucleated by the impurities 
out of the SP background, 
thus co-existing with SP in an inhomogeneous way. 
We note that although the amplitudes are modulated, 
the phases of both SP and AF order parameters 
are unchanged throughout the chain~\cite{Fukuyama,Saito}, 
i.e., there is no solitonic domain wall structure 
as discussed in Refs.~\cite{Khomskii,Fabrizio,Poilblanc}. 
The co-existence only survives at intermediate-$T$ 
and finally the uniform AF state is stabilized at low-$T$ 
where $u_i$ becomes zero. 
This co-existent state is similar to 
what is observed in slightly doped 
CuGeO$_3$ at low-$T$~\cite{Uchinokura,Fukuyama} 
(see Fig.~\ref{analogy}(a)). 

\begin{figure}
\epsfxsize=6.5cm
\centerline{\epsfbox{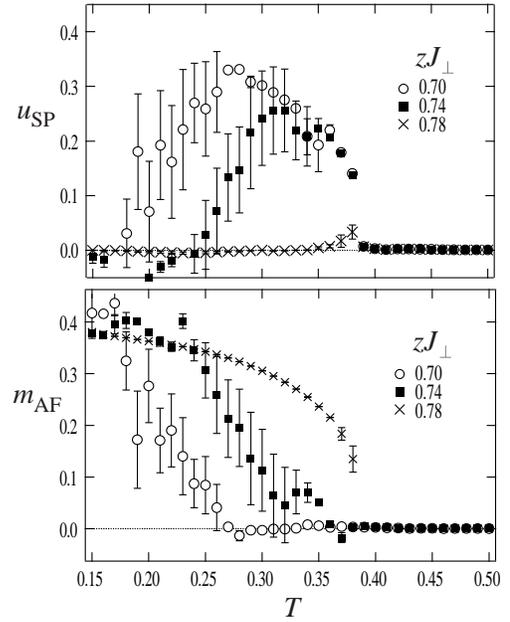}}
\caption{Temperature dependence of 
$u_{\rm SP}$ and $m_{\rm AF}$ 
for different values of $zJ_{\perp}$. 
Error bars are due to distribution of results from samples 
with different impurity positions. 
}
\label{uspandsz}
\end{figure}
In Fig.~\ref{uspandsz}, 
the infinite-$L$ extrapolation 
of the sample averaged SP lattice dimerization and AF spin moment, 
$u_{\rm SP}= \langle~\frac{1}{L}\left| \sum_i (-1)^i u_i \right|~\rangle_{\rm av}$ and 
$m_{\rm AF}= \langle~~\frac{1}{L}\left| \sum_i (-1)^i 
\langle S^z_i \rangle \right|~\rangle_{\rm av}$, 
respectively, 
are plotted as a function of $T$ for several values of $zJ_{\perp}$. 
In the region of $zJ_{\perp}$ shown in Fig.~\ref{uspandsz}, 
first $u_{\rm SP}$ develops as $T$ is lowered 
where the inhomogeneous 
SP state emerges, while $m_{\rm AF}=0$, 
as in the sample in Fig.~\ref{realspace}. 
Eventually $m_{\rm AF}$ becomes finite 
and both orders are stabilized as LRO. 
There, as $m_{\rm AF}$ develops, $u_{\rm SP}$ is suppressed, 
and finally the uniform AF state dominates over the SP state. 
Although the error bars due to random sample average are rather large, 
we can clearly see the re-entrant behavior in 
$u_{\rm SP}$, the region with co-existence, 
and the stability of the uniform AF state at low temperatures, 
to draw the phase diagram in Fig. \ref{phasediagram}. 
The slopes of 
the phase boundaries of the impurity 
induced uniform AF state and the co-existent state 
are considerably steep, 
therefore the states can be shifted drastically 
with small change of the interchain coupling. 

\begin{figure}
\epsfxsize=6.4cm
\centerline{\epsfbox{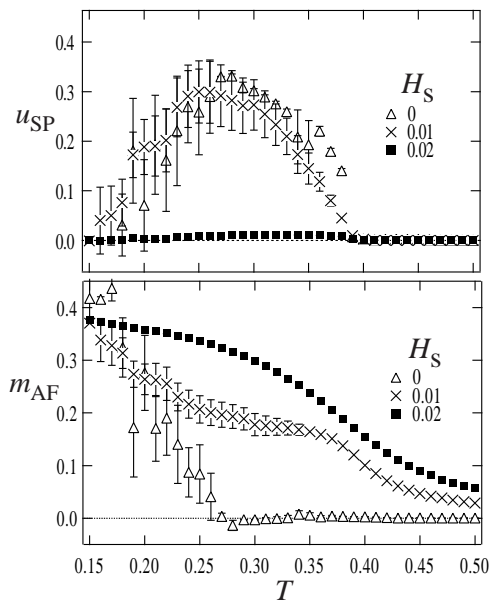}}
\caption{Temperature dependence of 
$u_{\rm SP}$ and $m_{\rm AF}$ for $zJ_{\perp}$=0.7 
under small staggered magnetic field, $H_{\rm S}$. 
}
\label{staggered}
\end{figure}
Next we apply staggered magnetic field $H_{\rm s}$
by adding the term, 
$H_{\rm s} \sum_i (-1)^i S^z_i$, 
to the effective 1D model in Eq.~(\ref{H_1D}). 
We consider this term to control the relative energies of SP and AF states,
analogous to the uniform magnetic field in CMR manganites.
In Fig.~\ref{staggered}, 
$T$-dependences of 
$u_{\rm SP}$ and $m_{\rm AF}$ are plotted for $zJ_{\perp}=0.7$, 
when the staggered magnetic field is varied. 
With considerably small values of $H_{\rm s}$, 
$u_{\rm SP}$ becomes completely suppressed, 
and at the same time $m_{\rm AF}$ is drastically increased. 
The inflection point in the $T$-dependence of $m_{\rm AF}$, 
representing the characteristic temperature 
for the emergence of AF, 
is shifted about $\Delta T = 0.2$ to higher-$T$, 
which is ten times the energy of the applied staggered magnetic field 
of $H_{\rm s}=0.02$. 
Thus we have demonstrated that 
there is actually a ``colossal'' response in this system. 

Now we compare our results to real systems. 
In doped CuGeO$_3$, 
in some range of $T$ above the impurity induced uniform AF phase, 
the static SP lattice distortions show 
a re-entrant behavior 
observed by elastic neutron scattering~\cite{Nakao} 
as mentioned previously, similar to what is seen in our calculations 
(Fig.~\ref{uspandsz}). 
However in experiments the correlation length remains finite~\cite{Nakao}, 
namely, the SP state remains to be SRO, 
while in our calculations it attains LRO. 
This discrepancy may be due to 
the treatment of the impurities in our calculations, 
such as to ignore effects of 
the random impurity positions between neighboring chains 
and/or to assume that the impurity modifies only $K_i$ neglecting 
the change in the intrachain coupling $J$. 
Such effects would make the system more disturbed 
resulting in frustration for the SP state 
in the interchain directions as pointed out in Ref.~\cite{Wang}, 
due to solitons of SP order parameters 
changing the phase at the impurities~\cite{Khomskii,Fabrizio,Poilblanc}. 
This actually reduces the correlation length of SP order parameter, 
but the SP lattice distortions would remain static 
as in the experiments even if such effects were considered. 
Therefore, we believe that main features of our results 
are relevant to doped CuGeO$_3$. 

We note that another theoretical study in Ref.~\cite{Mostovoy} also
showed apparently similar re-entrant behavior in a narrow
parameter range. However, because of their weak-coupling
continuum treatment with white noise-type disorder,
all of their states are with spatially uniform SP and AF order parameters. 
This is very different from the experiments~\cite{Uchinokura,Regnault,Nakao}, 
as well as from our SP and co-existent states, 
where we believe that the spatially inhomogeniety is crucial
for the emergence of the huge response.

Our observation that very weak perturbation to 
the system can drive such 
``colossal'' response should be relevant to the actual compounds. 
Although the staggered magnetic field that we have applied 
is a fictitious field, 
perturbations that can change the relative 
energies of SP and AF states might also produce such 
drastic changes. 
For example, 
the application of pressure, $p$, is one candidate, 
through the change in $J_{\perp}/J$ or $K_i$ 
resulting in a large change of the transition temperatures 
as we have seen in the phase diagram, Fig.~\ref{phasediagram}. 
In doped CuGeO$_3$, 
the hydrostatic pressure drives the system 
toward the stability of SP state, 
contrary to the usual expectations~\cite{SPreview}, 
due to the enhancement of 
the next-nearest-neighbor exchange along the chains~\cite{Uchinokura,pressure}. 
The enhanced $dT_{\rm c}/dp$ in doped samples 
compared with the pure sample may be related to the sensitivities 
of our impurity induced states. 
Furthermore, an estimate of uniaxial $dT_{\rm c}/dp$'s 
in terms of the Ehrenfest relation using 
the measured specific heat and thermal expansion~\cite{Lorenz} 
also gives a large value.  
The uniform magnetic field might 
lead to the giant response as well. 
Up to now, the experiments are mainly done for 
small-$x$ samples which found no such large change 
while the system converts into 
an incommensurate phase~\cite{Uchinokura,magnetic}. 
Measurements in the multicritical regime are desired. 

In conclusion, we have demonstrated
that a model of disordered SP system competing with
the AF order shows a re-entrant behavior and
huge responses to small perturbations
such as the staggered magnetic field or the change in
the interchain coupling $J_\perp$. 
This is analogous to the CMR effect in manganites,
where the multicritical phenomenon influenced by
the disorder leads to the ``colossal'' response.

We acknowledge Y. Tokura and Y. Tomioka 
for useful discussions and providing us data prior to publication, 
and M. Saito for critical readings of the manuscript. 


\begin{thebibliography}{99}
%
\bibitem{Hase}M. Hase {\it et al.}, Phys. Rev. Lett. {\bf 71}, 4059 (1993). 
\bibitem{Uchinokura}For reviews, K. Uchinokura, 
	J. Phys.: Condens. Matter {\bf 14}, R195 (2002); 
	T. Masuda, N. Koide, and K. Uchinokura, 
	Prog. Theor. Phys. Suppl. {\bf 145}, 306 (2002). 
\bibitem{Regnault}L. P. Regnault {\it et al.}, 
	Europhys. Lett. {\bf 32}, 579 (1995). 
\bibitem{Fukuyama}H. Fukuyama, T. Tanimoto, and M. Saito,
	J. Phys. Soc. Jpn. {\bf 65}, 1182 (1996). 
\bibitem{Khomskii}D. Khomskii, W. Geertsma, and M. Mostovoy, 
	Czech. J. Phys. {\bf 46}, Suppl. 6 3239 (1996). 
\bibitem{Fabrizio}M. Fabrizio and R. M{\'e}lin, 
	Phys. Rev. Lett. {\bf 78}, 3382 (1997). 
\bibitem{Mostovoy}M. Mostovoy, D. Khomskii, and J. Knoester, 
	Phys. Rev. B {\bf 58}, 8190 (1998). 
\bibitem{Saito}M. Saito, J. Phys. Soc. Jpn. {\bf 67}, 2477 (1998); 
	{\it ibid.} {\bf 68}, 2898 (1999); 
	M. Saito and M. Ogata, {\it ibid.} {\bf 71}, 721 (2002). 
\bibitem{Poilblanc}A. Dobry {\it et al.}, Phys. Rev. B {\bf 60}, 4065 (1999); 
	D. Augier, J. Riera, and D. Poilblanc, {\it ibid.} {\bf 61}, 
	6741 (2000).
\bibitem{CMRreview}Y. Tokura and N. Nagaosa, 
	Science {\bf 288}, 462 (2000). 
\bibitem{TomiokaPCMO}Y. Tomioka and Y. Tokura, 
	Phys. Rev. B {\bf 66}, 104416 (2002). 
\bibitem{Murakami}S. Murakami and N. Nagaosa, 
	Phys. Rev. Lett. {\bf 90}, 197201 (2003). 
\bibitem{Mn_disorder}D. Akahoshi {\it et al.}, 
	Phys. Rev. Lett. {\bf 90}, 177203
	(2003), and references therein. 
\bibitem{Tomioka55}Y. Tomioka {\it et al.},
	Phys. Rev. B {\bf 68}, 094417 (2003); 
	Y. Tomioka and Y. Tokura, unpublished. 
\bibitem{note_glassy}In stronger disorder case, an insulating
	spin-orbital-glass-like state is observed~\cite{Tomioka55}.
\bibitem{Motome}Y. Motome, N. Furukawa, and N. Nagaosa, 
	Phys. Rev. Lett. {\bf 91}, 167204 (2003). 
\bibitem{Aliaga}H. Aliaga {\it et al.}, Phys. Rev. B {\bf 68}, 104405 (2003).  
\bibitem{Nakao}H. Nakao {\it et al.}, 
	J. Phys. Soc. Jpn. {\bf 68}, 3662 (1999); 
	V. Kiryukhin  {\it et al.},
	Phys. Rev. B {\bf 61}, 9527 (2000). 
\bibitem{SPreview}J. W. Bray {\it et al.},  
	in {\it Extended Linear Chain Compounds} 
	edited by J. S. Miller 
	(Plunum Press, New York and London, 1983), p.353. 
\bibitem{Inagaki}S. Inagaki and H. Fukuyama, 
	J. Phys. Soc. Jpn. {\bf 52}, 3620 (1983). 
\bibitem{Imry}Y.Imry, P. Pincus, and D. Scalapino, Phys. Rev. B {\bf 12}, 
	1978 (1975); H. J. Schulz, Phys. Rev. Lett. {\bf 77}, 2790 (1996). 
\bibitem{Betsuyaku}H. Betsuyaku, 
	Prog. Theor. Phys. {\bf 73}, 319 (1985). 
\bibitem{note}For the numerical calculations, 
	we apply extrapolations in Trotter numbers $m=4-10$ 
	to infinite $m$ by polyminals of $1/m^2$. 
\bibitem{Wang}Y. J. Wang {\it et al.}, Phys. Rev. Lett. {\bf 83},
	1676 (1999). 
\bibitem{pressure}T. Masuda {\it et al.}, Phys. Rev. B {\bf 67}, 024423
	(2003. 
\bibitem{Lorenz}T. Lorenz {\it et al.}, Phys. Rev. B {\bf 56}, 501 (1997). 
\bibitem{magnetic}B. B{\" u}chner {\it et al.}, Phys. Rev. B {\bf 59}, 6886
	(1999). 
%
\end{thebibliography}
\end{document}